\documentclass{aastex}%
\usepackage{amsmath}
\usepackage{amsfonts}
\usepackage{amssymb}
\usepackage{graphicx}%
\begin{document}
\bigskip

{\Large Nonlinear plasma density modification by the ponderomotive force of
ULF pulsations at the dayside magnetosphere}

{\large \bigskip\ \ }

\ \ \ \ \ \ \ \ \ \ \ \ \ \ \ \ \ \ \ \ \ \ \ \ \ \ A. K. Nekrasov* and F. Z. Feygin**

\begin{center}
Institute of Physics of the Earth, Russian Academy of Sciences, 123995 Moscow, Russia

*) anekrasov@ifz.ru, **) feygin@ifz.ru

\bigskip
\end{center}

\textbf{Abstract} We investigate analytically and numerically a nonlinear
modification of the magnetospheric plasma density under the action of the
ponderomotive force induced by ULF traveling waves, using the nonlinear
stationary force balance equation. This equation is applied to both the dipole
and dayside magnetosphere having one and two minima of the geomagnetic field
near the magnetospheric boundary. The separate and joint actions of the
ponderomotive, centrifugal, and gravitational forces on the density
distribution are shown.


\bigskip

\textbf{1 Introduction}

\bigskip

The ultra-low frequency (ULF) electromagnetic pulsations ranging from tenths
of Hz up to a few Hz are a common phenomenon in the Earth's magnetosphere.
These waves are electromagnetic in character, i.e., have the electric and
magnetic components. This is confirmed by both ground measurements (e.g.
Troitskaya 1961; Tepley 1961; Heacock and Akasofu 1973; Bortnik et al. 2008)
and in-situ observations by satellites (e.g. Young et al. 1981; Erlandson et
al. 1990; Halford et al. 2010; Kim et al. 2010). The ULF pulsations are duct
along geomagnetic field lines in the direction of which the parameters of
magnetosphere are nonuniform. This inhomogeneity must be taken into account
when considering the influence of the wave action on the magnetospheric plasma
to obtain a more realistic picture of physical processes.

The theory of electromagnetic ion cyclotron waves in the ULF range has two
aspects. The first one is the problem of sources of these waves related to the
physics of the magnetosphere. This aspect is important not only for explaining
the origin of these waves and mechanisms by which they are generated, but also
for processes being responsible for the acceleration, diffusion and
precipitation of magnetospheric particles. These problems have got a great
deal of attention in past years (see e.g. Kennel and Petschek 1966; Cornwall
1966; Troitskaya and Guglielmi 1967; Feygin and Yakimenko,1971; Gendrin et al.
1971; Tverskoy 1971; Nekrasov 1987; Guglielmi and Pokhotelov 1996). Another
important aspect of the theory is the investigation of the influence of
ULF\ pulsations on the background space plasma. Ponderomotive forces induced
by these waves can significantly contribute to the plasma balance in the
Earth's magnetosphere (Allan 1992; Guglielmi et al. 1993; Guglielmi and
Pokhotelov 1994; Witt et al. 1995; Allan and Manuel 1996; Pokhotelov et al.
1996; Feygin et al. 1998; Nekrasov and Feygin 2005). In the papers by
Guglielmi et al. (1993, 1995), it was demonstrated for the dipole
magnetosphere that a pronounced maximum of plasma density is formed in the
vicinity of the minimum magnetic field force along the wave trajectory, if the
amplitude of electromagnetic ion cyclotron waves exceeds a certain critical
value. Near the dayside boundary of the magnetosphere, the geometry of
geomagnetic field lines changes from the structure with only one magnetic
field force minimum to the structure with two minima (so-called magnetic
holes)\ along the field line (Antonova and Shabansky 1968). Pokhotelov et al.
(1996) have shown that the ponderomotive force can lead to plasma density
accumulation in the vicinity of these magnetic holes.

In some papers cited above (e.g. Guglielmi et al. 1993, 1995), it has been
assumed that the equation of the force balance along the magnetic field lines
contains the total thermal pressure together with the nonlinear ponderomotive
force derived for small perturbations. However, it is not correct from the
point of view of the nonlinear theory, in which all terms must have the same
order of magnitude in one equation (for the correct approach see e.g. Allan
and Manuel 1996; Nekrasov and Feygin 2005, 2011). Therefore, conclusions made
in the corresponding papers are doubtful.

In this paper, we consider analytically and numerically the nonlinear
modification of the magnetospheric plasma density, using the nonlinear
stationary force balance equation. We apply our investigation to the dipole
and dayside magnetosphere having one and two minima of the geomagnetic field
near the magnetospheric boundary, respectively.

The paper is organized as follows. In Sect. 2, we give the expression for the
ponderomotive force of low-frequency waves traveling along the nonuniform
geomagnetic field. The geomagnetic field at the dayside of the magnetosphere
is discussed in Sect. 3. The stationary force balance equation is considered
in Sect. 4. In Sect. 5, numerical calculations of balance equation in the
curved geomagnetic field are given. Conclusive remarks are summarized in Sect. 6.

\bigskip

\bigskip\textbf{2 Ponderomotive force}

In this paper, we use an expression for the ponderomotive force
$F_{p\shortparallel}$ along the background nonuniform magnetic field induced
by the circularly-polarized electromagnetic WKB-waves traveling along field
lines (see e.g. Nekrasov and Feygin 2005). In the case $N^{2}>>1$, this
expression has the form$\ \ $%
\begin{equation}
F_{p\shortparallel}=-\frac{N^{2}E_{1}^{2}}{16\pi}\left\{  \nabla
_{\shortparallel}\ln\rho+\left[  \frac{\sigma\omega}{\omega_{i}-\sigma\omega
}+\left(  \frac{2\omega_{i}}{\omega_{i}-\sigma\omega}\right)  \frac{1}{N^{2}%
}\right]  \nabla_{\shortparallel}\ln B\right\}  ,
\end{equation}
where%
\begin{equation}
N^{2}=\frac{k_{\shortparallel}^{2}c^{2}}{\omega^{2}}\approx\frac{\omega
_{pi}^{2}}{\omega_{i}\left(  \omega_{i}-\sigma\omega\right)  }%
\end{equation}
is the refractive index. Here, $\omega_{pi}=\left(  4\pi ne^{2}/m_{i}\right)
^{1/2}$ and $\omega_{i}=eB/m_{i}c$ are the ion plasma and cyclotron
frequencies, respectively, $\rho=m_{i}n$ is the equilibrium plasma mass
density, $n$ is the local number density, $B$ is the local background magnetic
field, $E_{1}$ is the wave amplitude, $\omega$ ($>0$) is the wave frequency,
$k_{\shortparallel}$ is the wave number, $e$ and $m_{i}$ are the charge and
mass of the ions, $\sigma$ denotes the left-($+1$) and right-($-1$)
polarization, and $c$ is the speed of light in vacuum.

In the WKB approximation, we have $E_{1}\sim N^{-1/2}$ or $E_{1}^{2}%
N=E_{10}^{2}N_{0}=const,$ where the subscript $0$ here and below relates to
the corresponding values at the magnetic equator. We will express the
ponderomotive force through the amplitude of the wave magnetic field $B_{10}$
at the magnetic equator because satellite measurements give usually the
magnitude of magnetic field perturbations. From Faraday's equation for the
traveling wave, it is followed that $B_{1}=NE_{1}$. Thus, we obtain from these
relations and (2)%
\begin{equation}
N^{2}E_{1}^{2}=B_{10}^{2}\frac{N}{N_{0}}=B_{10}^{2}\left(  \frac{\rho}%
{\rho_{0}}\right)  ^{1/2}\frac{B_{0}}{B}\frac{\left(  1-\nu_{0}\right)
^{1/2}}{\left(  1-\nu_{0}\frac{B_{0}}{B}\right)  ^{1/2}},
\end{equation}
where $\nu_{0}=\sigma\omega/\omega_{i0}$. Substituting (3) in (1), we obtain%
\begin{align}
F_{p\shortparallel} &  =-\frac{B_{10}^{2}}{16\pi}\left(  \frac{\rho}{\rho_{0}%
}\right)  ^{1/2}\frac{B_{0}}{B}\frac{\left(  1-\nu_{0}\right)  ^{1/2}}{\left(
1-\nu_{0}\frac{B_{0}}{B}\right)  ^{1/2}}\\
&  \times\left\{  \nabla_{\shortparallel}\ln\rho+\frac{1}{1-\nu_{0}\frac
{B_{0}}{B}}\left[  \nu_{0}\frac{B_{0}}{B}+\frac{2}{N_{0}^{2}}\frac{\left(
1-\nu_{0}\frac{B_{0}}{B}\right)  }{\left(  1-\nu_{0}\right)  }\frac{\rho_{0}%
}{\rho}\frac{B^{2}}{B_{0}^{2}}\right]  \nabla_{\shortparallel}\ln B\right\}
.\nonumber
\end{align}

\bigskip

\textbf{3 Curved dayside geomagnetic field, field line equation and
longitudinal gradient }$\nabla_{\shortparallel}$\textbf{\ }

\bigskip

The magnetic data collected by satellites and theoretical models show a
complex structure of the geomagnetic field at the midday boundary of the
Earth's magnetosphere. This structure is characterized by a smooth transition
from "V" to a "W" magnetic field shape with two minima in the magnetic field
along field lines. There exist several analytical (Hones 1963; Mead 1964;
Antonova and Shabansky 1968) and quantitative models (Mead and Fairfield 1975;
Tsyganenko 1987) of the magnetic field in the Earth's magnetosphere based on
satellite observations. The model by Antonova and Shabansky (1968) is quite
convenient for an analysis of geophysical phenomena in the dayside
magnetosphere and, in addition, is in a reasonable agreement with the
magnetometer data provided by the HEOS 1, 2 satellites (Antonova et al. 1983).
In this model, the structure of the geomagnetic field is determined by the
existence of two dipoles: an original dipole having the Earth's magnetic
moment $M$ and an additional one which imitates the distortion of the magnetic
field due to the solar wind pressure. The latter dipole has the magnetic
moment $kM$, where $k$ is a constant parameter, and is shifted a distance $a$
(measured in the units of the Earth's radius $R_{E}$) at the dayside along the
Earth-Sun line from the position of the original dipole.

In the spherical coordinate system, magnetic field components for the
two-dipole model in the meridional noon-midnight plane have the form (Antonova
and Shabansky 1968)%
\begin{equation}
\ B_{r}=-\frac{2B_{E}x}{r^{3}}\alpha,
\end{equation}%
\begin{equation}
B_{\varphi}=\frac{B_{E}}{r^{3}}\sqrt{1-x^{2}}\beta,
\end{equation}
where $x=\sin\varphi$, $\varphi$ is the geomagnetic latitude, $r$ is measured
in units $R_{E}$, and $B_{E}$ $=0.311$ G is the equatorial magnetic field at
the Earth's surface. Coefficients $\alpha$ and $\beta$ are the following
(Antonova and Shabansky 1968):%
\begin{equation}
\ \alpha=1-\frac{kr^{3}\left(  a^{2}-2r^{2}+ar\sqrt{1-x^{2}}\right)
}{2\left(  a^{2}+r^{2}-2ar\sqrt{1-x^{2}}\right)  ^{5/2}},
\end{equation}%
\begin{equation}
\ \ \beta=1+\frac{kr^{3}\left[  \sqrt{1-x^{2}}\left(  a^{2}+r^{2}\right)
-ar\left(  2+x^{2}\right)  \right]  }{\sqrt{1-x^{2}}\left(  a^{2}%
+r^{2}-2ar\sqrt{1-x^{2}}\right)  ^{5/2}}.
\end{equation}
The value of the magnetic field $B$ in an arbitrary point of the field
line\ can be defined from (5) and (6) as%
\begin{equation}
B=\frac{B_{E}}{r^{3}}\left[  4x^{2}\alpha^{2}+\left(  1-x^{2}\right)
\beta^{2}\right]  ^{1/2}.
\end{equation}
An equation for the field line is determined by $dr/rd\varphi=B_{r}%
/B_{\varphi}$ or
\begin{equation}
\frac{dr}{dx}=-\frac{2xr}{1-x^{2}}\frac{\alpha}{\beta}.
\end{equation}

The intensity of the magnetic field force along the near boundary field lines
described by (9) can have two minima located symmetrically relative to the
magnetic equator (Antonova and Shabansky 1968). When $a$ tends to infinity,
the values $\alpha$ and $\beta$ (see 7 and 8) approach to $1$. In this case,
we have a transition to the one-dipole approximation. Thus, all the
expressions obtained in this paper can be transformed to those corresponding
to the one-dipole model. Since a purpose of our paper is a qualitative
description of the ponderomotive force action induced by geomagnetic
pulsations in the dayside magnetosphere along field lines at different
distances from the Earth (up to the magnetopause), we will not focus on
details of the two-dipole model.

The operator $\nabla_{\shortparallel}$ is defined by relation $\nabla
_{\shortparallel}=\mathbf{b}\cdot\mathbf{\nabla}$, where $\mathbf{b}$ is the
unit vector along the magnetic field. Using (5), (6), (9) and (10), we find%
\begin{equation}
\nabla_{\shortparallel}=2R_{E}^{-1}\eta^{-1/2}\frac{d}{dx},
\end{equation}
where%

\begin{equation}
\eta=\left(  \frac{dr}{dx}\right)  ^{2}+\frac{r^{2}}{1-x^{2}}.
\end{equation}

In order to investigate the ponderomotive force action induced by geomagnetic
pulsations in the dayside magnetosphere along field lines and carry out
analytical and numerical calculations at different distances from the Earth
(up to the magnetopause), we use the magnetic field model by Antonova and
Shabansky (1968) described above. For the equilibrium plasma mass density, we
take the power law form to describe the longitudinal field line distribution,
$\rho\sim r^{-\gamma}$. For the large distances from the Earth's surface which
we consider below, the best choice for $\gamma$ to be appropriate to
experimental data is $\gamma=1$ (Denton et al. 2006). Thus, we take
approximately
\begin{equation}
\rho\left(  x\right)  =\rho_{0}\left(  1-x^{2}\right)  ^{-1}.
\end{equation}
This formula can be applied up to $\varphi\approx\pm50-60^{0}$ (Denton et al.
2006). For simplicity, we do not take into account in (13) the non-dipole form
of the magnetic field (see 10) because of an uncertainty of plasma mass
distribution in this region.

\bigskip

\textbf{4 Stationary force balance equation}

\bigskip

From equations of motion for the ions and electrons in the second
approximation on the wave amplitude averaged over fast oscillations, we obtain
the force balance equation in the stationary state along the magnetic field
line (e.g. Nekrasov and Feygin 2005). Taking into account the gravitational
and centrifugal forces (e.g. Lemaire 1989; Persoon et al. 2009), we have
\begin{equation}
\nabla_{\shortparallel}p_{2}-\left(  g_{\shortparallel}+n_{\Omega
\shortparallel}\Omega^{2}R_{E}r\cos^{2}\varphi\right)  \rho_{2}%
=F_{p\shortparallel}.
\end{equation}
Here $p_{2}=\rho_{2}c_{s}^{2}$ and $\rho_{2}$ are nonlinear stationary
perturbations of pressure and mass density, respectively, $c_{s}=\left(
2T/m_{i}\right)  ^{1/2}$ is the sound speed, $T$ is the temperature,
$g_{\shortparallel}=\mathbf{g\cdot b=-}g_{E}B_{r}/r^{2}B$ is the longitudinal
gravitational acceleration ($g_{E}=9.8$ m sec$^{-2}$), $\Omega$ is the Earth's
rotation frequency, and $n_{\Omega\shortparallel}=\mathbf{n}_{\Omega}%
\cdot\mathbf{b}$, where $\mathbf{n}_{\Omega}$ is the unit vector along the
centrifugal force. Calculations show that $n_{\Omega\shortparallel}=\left(
1+\beta/2\alpha\right)  \left(  B_{r}/B\right)  \cos\varphi$. Substituting (4)
into (14) and using (5), (11), and (13), we find
\begin{align}
\frac{d}{dx}\frac{\rho_{2}}{\rho_{0}} &  =-\frac{R_{E}}{2c_{s}^{2}r^{2}}%
\eta^{1/2}\frac{B_{r}}{B}\left[  g_{E}-\left(  1+\beta/2\alpha\right)
\Omega^{2}R_{E}r^{3}\cos^{3}\varphi\right]  \frac{\rho_{2}}{\rho_{0}}%
-\frac{B_{10}^{2}}{16\pi\rho_{0}c_{s}^{2}}\frac{B_{0}}{B}\frac{\left(
1-\nu_{0}\right)  ^{1/2}}{\left(  1-\nu_{0}\frac{B_{0}}{B}\right)  ^{1/2}}\\
&  \times\left\{  \frac{2x}{\left(  1-x^{2}\right)  ^{3/2}}+\left[  \frac
{1}{\left(  1-x^{2}\right)  ^{1/2}}\frac{\nu_{0}}{\left(  1-\nu_{0}\frac
{B_{0}}{B}\right)  }\frac{B_{0}}{B^{2}}+\frac{\left(  1-x^{2}\right)  ^{1/2}%
}{2\pi\rho_{0}c^{2}}B\right]  \frac{dB}{dx}\right\}  .\nonumber
\end{align}
This differential equation determines a nonlinear redistribution of the plasma
density due to the action of the ponderomotive force.

Equation (15) can be rewritten in the form%
\begin{equation}
\frac{d}{dx}\frac{\rho_{2}}{\rho_{0}}=A_{1}\frac{\rho_{2}}{\rho_{0}}%
+A_{2}\left(  A_{3}+A_{4}+A_{5}\right)  .
\end{equation}
Coefficients $A_{i}$, $i=1,2,3,4,5,$ are the following:%

\begin{align}
A_{1} &  =\frac{g_{eff}R_{E}x\alpha}{\left(  1-x^{2}\right)  rc_{s}^{2}\beta
},A_{2}=-\frac{B_{10}^{2}}{16\pi\rho_{0}c_{s}^{2}}\frac{B_{0}}{B}\frac{\left(
1-\nu_{0}\right)  ^{1/2}}{\left(  1-\nu_{0}\frac{B_{0}}{B}\right)  ^{1/2}%
},A_{3}=\frac{2x}{\left(  1-x^{2}\right)  ^{3/2}},\\
A_{4} &  =\frac{1}{\left(  1-x^{2}\right)  ^{1/2}}\frac{\nu_{0}}{\left(
1-\nu_{0}\frac{B_{0}}{B}\right)  }\frac{B_{0}}{B^{2}}\frac{dB}{dx},A_{5}%
=\frac{\left(  1-x^{2}\right)  ^{1/2}}{2\pi\rho_{0}c^{2}}B\frac{dB}%
{dx},\nonumber
\end{align}
where we have used (9) and (12) in the first term on the right-hand side of
(16). The acceleration $g_{eff}$ in the coefficient $A_{1}$ is equal to%
\begin{equation}
g_{eff}=g_{E}+g_{cf}=g_{E}-\left(  1+\beta/2\alpha\right)  \Omega^{2}%
R_{E}r^{3}\cos^{3}\varphi,
\end{equation}
where $g_{E}$ and $g_{cf}$ describe an influence of the gravitational and
centrifugal forces, respectively.

If we take for estimation $\alpha\sim\beta\sim1$ in (18) and substitute
numerical values of parameters $g_{E}$, $\Omega$, and $R_{E}$, we obtain
\begin{equation}
\frac{g_{E}}{g_{cf}}\sim\frac{3\times10^{2}}{r^{3}\cos^{3}\varphi},
\end{equation}
where $r\approx r_{0}\cos^{2}\varphi$ (see 10). Thus, for $r_{0}\cos
^{3}\varphi<6.7$, one can neglect the contribution of the centrifugal force.
For example, the last condition for $r_{0}=8$ is satisfied at $\left\vert
\varphi\right\vert \gtrsim20^{0}$. The same for $r_{0}=10$ is at $\left\vert
\varphi\right\vert \gtrsim30^{0}$. In the region of the magnetic equator, the
centrifugal force is dominant and leads to a local peak in mass density. An
estimation for $r_{0}=8$ corresponds to the results by Denton et al. (2006)
(see their Figs. 9 and 10).

The coefficient $A_{1}$ given by (17) will be the same in the equation for the
background equilibrium. Its estimation for gravity, for example, at $r_{0}=8$
is the following: $A_{1}\sim8\times10^{-3}x\left(  1-x^{2}\right)  ^{-2}$ (we
have taken $c_{s}^{2}=10^{9}$ m$^{2}$ sec$^{-2}$ for $T=10$ eV). Thus, one
could wait an essential contribution of the gravitational force to the
equilibrium mass density distribution at $1-x^{2}\lesssim4\times10^{-3}$.
However, this result does not coincide with (13). This could be a consequence
of additional factors except for the gravitational force which lead to the
power law distribution of mass density.

We see from (16) that the effective gravity $g_{eff}$ and ponderomotive force
$\sim A_{2}$ influence on the nonlinear density redistribution. These two
forces can be of the same order of magnitude or one force can dominate in the
dependence on the wave amplitude. We note also that the nonlinear disturbance
of density increases when $\nu_{0}B_{0}\rightarrow B$ or $\omega
\rightarrow\omega_{i}$. This case can occur in magnetic holes.

$\bigskip$

\textbf{5 Numerical results}

\bigskip

For numerical calculations, we have chosen $a=33$ and $k=13$. Such parameters
correspond to the dayside boundary of the magnetosphere at the distance
$10R_{E}$ and the region of experimental data obtained by HEOS 1, 2 satellites
(Antonova et al. 1983). Solution of Eq. (16) has been carried out by means of
the Runge-Kutta method. In all the cases considered, we assumed that the
square of the sound velocity $c_{s}^{2}=10^{9}$ m$^{2}$ sec$^{-2}$. As it has
been mentioned above, this value corresponds to $T=10$ eV. The plasma mass
density at the equator $\rho_{0}$ was chosen equal to $1.67\times10^{-20}$ kg
m$^{-3}$\ for all $L$ ($L$ is McIlwain's parameter) near the midday boundary
of the Earth's magnetosphere. These values are based on experimental data from
measurements of the plasma density from the plasmapause up to the
magnetosphere boundary. In these areas, the plasma density depends weakly on
$L$ (Chappel 1974; Carpenter and Anderson 1992). Since the purpose of this
paper is a qualitative study of the ponderomotive force effect of
low-frequency perturbations on the redistribution of the background plasma
density, we do not focus on details of the weak dependence of $\rho_{0}$ on
$L$.

The dependence of the Earth's magnetic field $B$ on the geomagnetic latitude
$\varphi$ in the meridional plane of the dayside magnetosphere for different
$L$ for the model by Antonova and Shabansky (1968) is given in Fig. 1. This
figure shows a smooth transition from the "V" to "W" magnetic field shape with
two minima in the magnetic field along the field lines. Fig. 2 demonstrates
the separate and joint influence of\ the gravitational, centrifugal, and
ponderomotive forces on the relative latitudinal density perturbation
$\delta=\rho_{2}/\rho_{0}$ for $L=6$ and $k=13$ (see 16 and 17). The same is
shown in Fig. 3 at $L=10$. We see that the role of gravity decreases and the
mass density is peaked due to ponderomotive and centrifugal forces. Such
peaking is observed in the magnetosphere of the Earth (Denton et al. 2006).
Figures 4 and 5 represent the difference in $\delta$ distribution for the
dipole ($k=0$) and non-dipole ($k=13$) geomagnetic field at $L=8$ and $L=10$,
respectively. We see that the essential difference is\ due to the presence of
geomagnetic holes. We note that small variations in the density distribution
given in Figs. 2-5 are because of the small value of the coefficient $A_{1}$
in (16). In order to show different dependence's, we have taken a small value
for $B_{10}$.

At the magnetic equator, the projection of the ponderomotive, gravitational,
and centrifugal forces on the geomagnetic field line is equal to zero. The
longitudinal centrifugal force $f_{cf\shortparallel}$ is expressed via the
longitudinal gravitational force $f_{g\shortparallel}$ at $\cos\varphi\sim1$
and for $r_{0}=8\div10$ in the form%
\[
f_{cf\shortparallel}\sim1.7\div3.3f_{g\shortparallel}%
\]
(see 19). Thus roughly, these forces are of the same order of magnitude and
small (see discussion below 19). Their influence is displayed when the
ponderomotive force is also small (Figs. 2-5). At the wave amplitudes $B_{10}$
larger than $10^{-6}$ G, the gravitational and centrifugal forces are not important.

\bigskip

\textbf{6 Conclusion}

\bigskip

In this paper, we have performed an analytical and numerical study of the
influence of the geomagnetic field structure near the dayside magnetospheric
boundary on the plasma density redistribution due to the ponderomotive force
action induced by ULF perturbations.\ We have shown that the ponderomotive
force causes the nonlinear stationary plasma density redistribution, which is
much smaller than the background plasma density. We note that the
ponderomotive force is a nonlinear effect, which is found by using linear
perturbations. The assumption that the ponderomotive force has the same order
of magnitude as the background pressure and gravity is incorrect. However,
such an assumption has been adopted in some papers (Guglielmi et al. 1993,
1995; Guglielmi and Pokhotelov 1994; Pokhotelov et al. 1996; Feygin et al.
1998). Ponderomotive force (1) is derived by the perturbation method.
Therefore, its contribution can only disturb an equilibrium state. It is
followed from (15) that $\rho_{2}/\rho_{0}\sim B_{10}^{2}/4\pi\rho_{0}%
c_{s}^{2}$ on the order of magnitude (see also Nekrasov and Feygin 2005). For
parameters given above and for $B_{10}=5\times10^{-6}$ G, the ratio $\rho
_{2}/\rho_{0}$ is of the order of $10^{-2}$.

We have investigated the plasma density redistribution for the model of the
geomagnetic field $B$ by Antonova and Shabansky (1968) (Fig. 1). This choice
is justified by its simplicity. The use of a more complex geometry for $B$ can
be useful for an analysis of certain particular situations (for example,
specific satellite measurements etc). The results, however, will be
qualitatively the same. We have shown that the two-dipole structure of the
Earth's magnetic field (geomagnetic holes) leads to an essential modification
in the plasma distribution (Figs. 4 and 5). In Figs. 2-5, we observe the
appearance of two disturbed plasma density minima under the action of ULF
pulsations (the curves 1 in Figs. 2 and 3 display the action of the
centrifugal and gravitational forces). This two minima effect could be
verified by observations.

\bigskip

\textbf{Acknowledgements}

The authors acknowledge the financial support from the Russian Foundation for
Basic Research, research grant No. 10-05-00376, and the Program of Russian
Academy of Sciences No. 4 and No. 22.

\bigskip

\textbf{References}

\bigskip

Allan, W.\textbf{:} J. Geophys. Res. \textbf{97}, 8483 (1992)

Allan, W., Manuel, J. R.: Ann. Geophys. \textbf{14}, 893 (1996)

Antonova, A.E., Shabansky, V.P.: Geomagn. Aeron. \textbf{8}, 639 (1968)

Antonova, A.E., Shabansky, V.P., Hedgecock, P.C.: Geomagn. Aeron. \textbf{23},
574 (1983)

Bortnik, J., Cutler, J.W., Dunson, C., Bleier, T.E., McPherron, R.L.: J.
Geophys. Res. \textbf{113}, A04201 (2008)

Carpenter, L.R., Anderson, R.R.:\ J. Geophys. Res. \textbf{97}, 1097 (1992)

Chappel, C.R.:\ J. Geophys. Res. \textbf{79}, 1861 (1974)

Cornwall, J.M.: J. Geophys. Res. \textbf{71}, 2185 (1966)

Denton, R.E., Takahashi, K., Galkin, I.A., Nsumei, P.A., Huang, X., Reinisch,
B.W., Anderson, R.R., Sleeper, M.K., Hughes, W.J.: J. Geophys. Res.
\textbf{111}, A04213 (2006)

Erlandson, R.E., Zanetti, L.J., Potemra, T.A.: J. Geophys. Res. \textbf{95},
5941 (1990)

Feygin, F.Z., Yakimenko, V.L.: Ann. Geophys. \textbf{27}, 49 (1971)

Feygin, F.Z., Pokhotelov, O.A., Pokhotelov, D.O., Mursula, K., Kangas, J.,
Braysy, T., Kerttula, R.: J. Geophys. Res. \textbf{103}, 20.481 (1998)

Gendrin, R., Lacourly, S., Roux, A., Solomon, J., Feygin, F.Z., Gokhberg,
M.B., Troitskaya, V.A., Yakimenko, V.L.: Planet. Space Sci. \textbf{19}, 165 (1971)

Guglielmi, A.V., Pokhotelov, O.A., Stenflo, L., Shukla, P.K.: Astrophys. Space
Sci. \textbf{200}, 91 (1993)

Guglielmi, A.V., Pokhotelov, O.A.: Space Sci. Rev. \textbf{65}, 5 (1994)

Guglielmi, A.V., Pokhotelov, O.A., Feygin, F.Z., Kurchashov, Yu.P., McKenzie,
J.F., Shukla, P.K., Stenflo, L., Potapov, A.S.: J. Geophys. Res. \textbf{100},
7997 (1995)

Guglielmi, A.V., Pokhotelov, O.A.: Geoelectromagnetic waves, IOP Publ. Ltd,
Bristol. (1996)

Halford, A.J., Fraser, B.J., Morley S.K.: J. Geophys. Res. \textbf{115},
A12248 (2010)

Heacock, R., Akasofu, S.: J. Geophys. Res. \textbf{78}, 5524 (1973)

Hones, E.F.: J. Geophys. Res. \textbf{68}, 1209 (1963)

Kennel, C.F., Petschek, H.E.: J. Geophys. Res. \textbf{71}, 1 (1966)

Kim, H., Lessard, M.R., Engebretson, M.J., L\"{u}hr, H.: J. Geophys. Res.
\textbf{115}, A09310 (2010)

Lemaire, J.: Phys. Fluids B \textbf{1}, 1519 (1989)

Mead, G.D.: J. Geophys. Res. \textbf{69}, 1181 (1964)

Mead, G.D., Fairfield, D.H.: J. Geophys. Res. \textbf{80}, 523 (1975)

Nekrasov, A.K.: Geomagn. Aeron. \textbf{27}, 705 (1987)

Nekrasov, A.K., Feygin, F.Z.: Phys. Scripta \textbf{71}, 310 (2005)

Nekrasov, A.K., Feygin, F.Z.: Nonlin. Proc. Geophys. \textbf{18}, 235 (2011)

Persoon, A.M., Gurnett, D.A., Santolik, O., Kurth, W.S., Faden, J.B., Groene,
J.B., Lewis, G.R., Coates, A.J., Wilson, R.J., Tokar, R.L., Wahlund, J.-E.,
Moncuquet, M.: J. Geophys. Res. \textbf{114}, A04211 (2009)

Pokhotelov, O.A., Feygin, F.Z., Stenflo, L., Shukla, P.K.: J. Geophys. Res.
\textbf{101}, 10.827 (1996)

Tepley, L.R.: J. Geophys. Res. \textbf{66}, 1651 (1961)

Troitskaya, V.A.: J. Geophys. Res. \textbf{66}, 5 (1961)

Troitskaya, V.A., Guglielmi, A. V.: Space Sci. Rev. \textbf{7}, 689 (1967)

Tsyganenko, N.A.: Planet. Space Sci. \textbf{35}, 1347 (1987)

Tverskoy, B.A.: Dynamics of the Earth's Radiation Belts, NASA,
Technctranslation, F-635 (1971)

Witt, E., Hudson, M.K., Li, X., Roth, I., Temerin, M.: J. Geophys. Res.
\textbf{100}, 12.151 (1995)

Young, D.T., Perraut, S., Roux, A., de Villedary, C., Gendrin, R., Korth, A.,
Kremser, G., Jones, D.: J. Geophys. Res. \textbf{86}, 6755 (1981)

\bigskip

\textbf{Figure captions}

\bigskip

Fig. 1. The dependence of the Earth's magnetic field $B$ on the geomagnetic
latitude $\varphi$ in the meridional plane of the dayside magnetosphere for
different $L$ in the model by Antonova and Shabansky (1968).

Fig. 2. The dependence of the relative density perturbation $\delta$ for
$L=6$, $k=13$. The curve $1$ corresponds to $g_{eff}\neq0$, $B_{10}=0$, curve
$2$ - $g_{eff}\neq0$, $B_{10}=10^{-6}$ G, and curve 3 - $g_{eff}=0$,
$B_{10}=10^{-6}$ G.

Fig. 3. The same as in Fig. 2 for $L=10$.

Fig. 4. The dependence of $\delta$ distribution on the dipole ($k=0$) and
non-dipole ($k=13$) geomagnetic field for $L=8$ ($g_{eff}\neq0$,
$B_{10}=10^{-6}$ G).

Fig. 5. The same as in Fig. 4 for $L=10$.

\bigskip

\bigskip

\end{document}